\documentclass[aip,jap,reprint]{revtex4-1}

\usepackage[pdftex]{graphicx}
\usepackage{amsmath}
\usepackage{amssymb}
\usepackage{bm}

\usepackage[utf8]{inputenc}
\usepackage[T1]{fontenc}
\usepackage{lmodern}

\usepackage[dvipsnames]{xcolor}
\usepackage[colorlinks, urlcolor=blue, citecolor=blue, linkcolor=blue, pdfstartview=FitH]{hyperref}
\usepackage[all]{hypcap}

\def\affiSOLAB{Spin\ Optics\ Laboratory, Saint~Petersburg\ State\ University, Peterhof, 198504 St.~Peterbsurg, Russia}
\def\affiE2{Experimentelle\ Physik\ 2, Technische\ Universit\"at\ Dortmund, D-44221 Dortmund, Germany}
\def\affiPhoton{Photonics\ Department, Saint~Petersburg\ State\ University, Peterhof, 198504 St.~Petersburg, Russia}

\begin{document}

\title{Stimulated spin noise in an activated crystal}

\author{M.~M.\ Sharipova}
\affiliation{\affiSOLAB}

\author{A.~N.\ Kamenskii}
\affiliation{\affiE2}

\author{I.~I.\ Ryzhov}
\affiliation{\affiPhoton}
\affiliation{\affiSOLAB}

\author{M.~Yu.\ Petrov}
\affiliation{\affiSOLAB}

\author{G.~G.\ Kozlov}
\affiliation{\affiSOLAB}

\author{A.\ Greilich}
\affiliation{\affiE2}

\author{M.\ Bayer}
\affiliation{\affiE2}

\author{V.~S.\ Zapasskii}
\affiliation{\affiSOLAB}

\date{\today}

\begin{abstract}
In the spin noise spectroscopy, the magnetic susceptibility spectrum is known to be provided by the spin-system untouched by any external perturbation, or, better to say, disturbed only by its thermal bath. 
We propose a new version of the spin noise spectroscopy, with the detected magnetization (Faraday-rotation) noise being \emph{stimulated} by an external fluctuating magnetic field with a quasi-white spectrum. 
Experimental study of the stimulated spin noise performed on a $\mathrm{BaF}_2 : \mathrm{U}^{3+}$ crystal in a longitudinal magnetic field has revealed specific features of this approach and allowed us to identify the Van-Vleck and population-related contributions to the AC susceptibility of the system and to discover unusual magnetic-field dependence of the longitudinal spin relaxation rate in low magnetic fields. 
It is shown that spectra of the stimulated and spontaneous spin noise, being both closely related to the spin-system magnetic susceptibility, are still essentially different. 
Distinctions between the two types of the spin-noise spectra and two approaches to the spin noise spectroscopy are discussed. 
\end{abstract}


\maketitle 

\section{Introduction}

The most important characteristics of paramagnetic substances are known to be contained in the spectra (frequency dependences) of their magnetic susceptibility. 
These spectra, depending on the frequency of the AC field and its alignment with respect to the external DC field, provide information about relaxation properties and energy structure of the system. 
In particular, the magnetic resonance technique intended to monitor spin-precession resonances investigates the magnetic susceptibility spectra in relatively high fields, when the magnetic splitting of the spin-system substantially exceeds the spin-relaxation rates. 
The measurements of low-frequency AC susceptibility that have been started and primarily developed by C.~Gorter~\cite{Gorter} were mainly performed in the range of low magnetic fields (down to zero) and were aimed at studying the spin-spin interactions and the relaxation-controlled response of the system. 
At present, the measurements of the `nonresonant' magnetic susceptibility are widely used for studying diverse magnetic systems, alloys, superconductors, and minerals~\cite{1984-AC,2003-AC,2003-ACJ,2003-ACP,2009-AC,2013-AC}. 

Experimentally, spectral features of the AC susceptibility are revealed either as alterations of the AC field itself or as changes in characteristics of the perturbed spin-system. 
The first case is realized, e.g., in the conventional EPR spectrometers~\cite{Alt}, when the resonant absorption of the AC field affects Q-value of the microwave cavity and, thus, changes the AC field inside the cavity. 
Among numerous methods of the second type~\cite{EPR-73, Zap-PP, EPR-80, EPR-80a, EPR-81, EPR-82}, an important place is occupied by optical methods, when the effect of the AC field is detected by optical means~\cite{Giltner, Zap-Mac, Zvezdin}. 

In the last decade, an alternative method of measuring the AC-susceptibility spectrum that did not use the AC field at all has emerged~\cite{Muller, Zap-rev, Hubner, Glazov}. 
This method, usually referred to as spin noise (SN) spectroscopy, implies measuring spontaneous fluctuations of the magnetization (Faraday rotation), which, in accordance with the fluctuation-dissipation theorem, should reproduce the spectrum of its linear susceptibility~\cite{Landau}. 
Since the SN spectroscopy, in contrast to traditional methods of linear response, generally does not imply application of any external AC field, it is considered to be nonperturbative. 
At present, this experimental technique, first developed for atomic systems~\cite{AlZap, Crooker}, is also widely applied to semiconductors, for which it proved to be highly efficient and informative~\cite{Muller, Hubner}. 

As has been suggested in Ref.\ \onlinecite{GIRI}, the applicability of the SN spectroscopy to a particular paramagnet can be estimated using the notion of the Faraday-rotation (FR) cross-section numerically equal to the FR angle in the medium of unit length with the unit spin density. 
In Ref.\ \onlinecite{FTT}, we have compared the values of this quantity for paramagnets of different classes (alkali-metal vapors, semiconductors, and dielectric crystals with paramagnetic impurities). 
It was found that for crystals with heavy paramagnetic ions this quantity is approximately $4$ orders of magnitude smaller than for semiconductor systems with favorable magneto-optical characteristics. 
As an example of an impurity of this kind, we used divalent ion of thulium, characterized by strongest magneto-optical activity among all the rare-earth ions. 
Thus, we came to the conclusion that involvement of crystals and glasses with transition-metal ions (iron group, lanthanides, and actinides) into the circle of objects of the SN spectroscopy would require a great deal of efforts. 

Notice that specific merits of the SN spectroscopy are not restricted to its nonperturbativity. 
In particular, when passing to SN measurements in semiconductors, whose FR cross-sections were several orders of magnitude smaller than in atomic systems~\cite{FTT}, the sensitivity of the measurements has been effectively increased by a few orders of magnitude by replacing the scanning spectrum analyzers with those of the Fourier-transform type~\cite{Romer, Crooker1}. 
It is also noteworthy that though the FR of heavy paramagnetic ions per unit spin is really small, absolute concentrations of the ions in materials of practical importance (laser media, optical switches, modulators, up-converters, etc.) are usually fairly high ($\sim 10^{19}$ cm$^{-3}$), and total magnitude of the paramagnetic FR, at low temperatures, appears to be high. 
In other words, the FR, in these systems, can respond highly sensitively to small variations of the external magnetic field. 

In this paper, we propose to exploit this high magneto-optical sensitivity of the activated crystals to realize a sort of SN spectroscopy with the detected SN being \emph{stimulated} rather than spontaneous. 
The proposed method implies detection of the FR noise spectrum of a paramagnet in an external \emph{fluctuating} magnetic field with a properly shaped spectrum of the fluctuations. 
This method (hereafter referred to as \emph{stimulated SN spectroscopy}) can be considered as a combination of the modulation magneto-optical technique~\cite{Zap-mod, ZKM-1} and conventional SN spectroscopy. 
We will show that spectra of the spontaneous and stimulated SN, being both directly connected with magnetic AC susceptibility of the system, are still not identical. 
In our opinion, the proposed experimental method is useful not only as a new approach to the SN spectroscopy of paramagnets with low FR cross-section but may also serve as an instructive illustration to the applicability of the fluctuation-dissipation theorem. 
To the best of our knowledge, we present here the first application of SN spectroscopy (though in a modified form) to dielectric crystals with paramagnetic ions. 

The	paper is organized as follows. 
In Sec.\ \ref{sec2}, we briefly consider the theoretical background of the problem of stimulated and spontaneous SN and present basic equations that describe the relation between spectra of the AC~susceptibility and of the spontaneous and stimulated SN. 
In Sec.\ \ref{sec3}, we outline characteristics of the experimental setup intended for detecting the stimulated SN and describe some expedients used to increase the sensitivity of the measurements. 
Results of the measurements are presented in Sec.\ \ref{sec4} and discussed in more detail in Sec.\ \ref{sec5}. 
A few concluding remarks are given in Sec.\ \ref{sec6}. 

\section{\label{sec2}Theoretical Background}

Let us consider distinctive properties of the spontaneous and stimulated SN detected, respectively, in the absence of any external perturbation and in the presence of a randomly fluctuating magnetic field. 
We consider the simplest case of a two-level spin system ($S=1/2$) in a static magnetic field at high temperatures (with thermal energy $kT$ strongly exceeding the Zeeman energy). 
We will be interested in the magnetic susceptibility $\chi_ \omega$ of the system that connects amplitude of a small AC magnetic field $h_ \omega e^{-\imath\omega t}$ with that of the induced magnetization $m_\omega e^{-\imath\omega t}$: $m_\omega = \chi_\omega h_\omega$. 
In addition, we will assume that, in the laboratory coordinate system, the small AC magnetic field has only $z$ component, while the applied static magnetic field has both $z$ and $x$ components: ${\bf H} = (H_x, 0, H_z)$. 
The Hamiltonian of such a system has the form:
\begin{equation}
	H=g\beta [H_z \hat S_z+H_x\hat S_x +h_\omega e^{-\imath\omega t} \hat S_z].
	\label{1}
\end{equation}
 
The calculation~\cite{ZKM} based on the Bloch equations with the relaxation parameters $T_1$ and $T_2$ (the longitudinal and transverse relaxation times) shows that susceptibility of this simplest system is given by the expression
\begin{equation}
	\chi_\omega =\chi_{st}\left[ {\omega^2_0 \cos^2\alpha \over \omega^2_0-\omega ^2-2\imath \omega /T_2+1/T_2^2 }+{ \sin^2\alpha \over 1-\imath\omega T_1}\right]
\end{equation}
where 
\begin{equation*}
\mathop{\rm tg}\alpha ={H_z\over H_x}, \;
\omega_0 \equiv {g\beta\over \hbar}\sqrt {H_x^2+H_z^2}, \;
\chi_{st}= {g^2\beta^2S(S+1)\over k_BT},
\end{equation*}
(notations are conventional).

One can see from this expression that the susceptibility, in the range of low frequencies ($\omega\ll \omega_0,T_2^{-1}$), contains two contributions, one of them, usually referred to as of Van-Vleck type or adiabatic ($\chi_{VV}$), being inertialess, while the second---the population-related contribution ($\chi_P$)---is characterized by the response time $T_1$ \cite{ZKM}:
\begin{equation}
	\chi_{VV}=\chi_{st} {H_x^2\over H_x^2+H_z^2},\quad
 	\chi_P={\chi_{st} \over 1-\imath\omega T_1}{H_z^2\over H_x^2+H_z^2}.
\label{3}
\end{equation}

The technique of stimulated SN spectroscopy, as has been mentioned above, includes detection of the magnetization noise $m(t)$ induced by the `white' fluctuating magnetic field $h(t), (\langle h(t)h(t')\rangle =V\delta (t-t')$ with $V$ describing the ``strength'' of the fluctuating magnetic field), and subsequent Fourier transformation of its correlation function $K(t_1-t_2)\equiv \langle m(t_1)m(t_2)\rangle$. 
The function thus obtained ${\cal M}(\omega)=\int e^{\imath \omega t}K(t) dt $ reflects frequency dependence of the magnetic AC susceptibility of the system. 
It can be shown that 
\begin{equation}
	{\cal M}(\omega) = V|\chi_\omega|^2.
\end{equation}
The noise power spectrum ${\cal N}(\omega)$ observed in the spontaneous SN spectroscopy, for the considered spin system, is given by the expression 
\begin{align}
	{\cal N}(\omega)=&{\cos^2\alpha\over 8\pi T_2} \, \left[
	{1\over 1/T_2^2+[\omega-\omega_0]^2}+{1\over 1/T_2^2+[\omega+\omega_0]^2}
	\right] \nonumber\\
					&+\hskip1mm{\sin^2\alpha\over 4\pi T_1}\hskip1mm{1\over 1/T_1^2+\omega^2}.
	\label{5}
\end{align}
 
One can make sure that, in accordance with the fluctuation-dissipation theorem, ${\cal N}(\omega)\sim \mathop{\rm Im}\chi_\omega/\omega$.

The functions ${\cal N}(\omega)$ and ${\cal M}(\omega)$ detected, respectively, in the spectroscopy of spontaneous and stimulated SN, as is seen, exhibit similar spectral behavior (peaks at $\omega=\pm\omega_0$ and $\omega=0$). 
Moreover, it seems correctly to say that in the stimulated SN spectroscopy the detected signal is a response to stochastic perturbation created by the experimentalist, whereas in the spectroscopy of spontaneous SN this stochastic perturbation is created by the thermal bath of the spin system. 
Still, the functions ${\cal N}(\omega)$ and ${\cal M}(\omega)$ appear to be different. 
Consider it in more detail. 
 
\begin{figure}[t]
\includegraphics[width=\columnwidth]{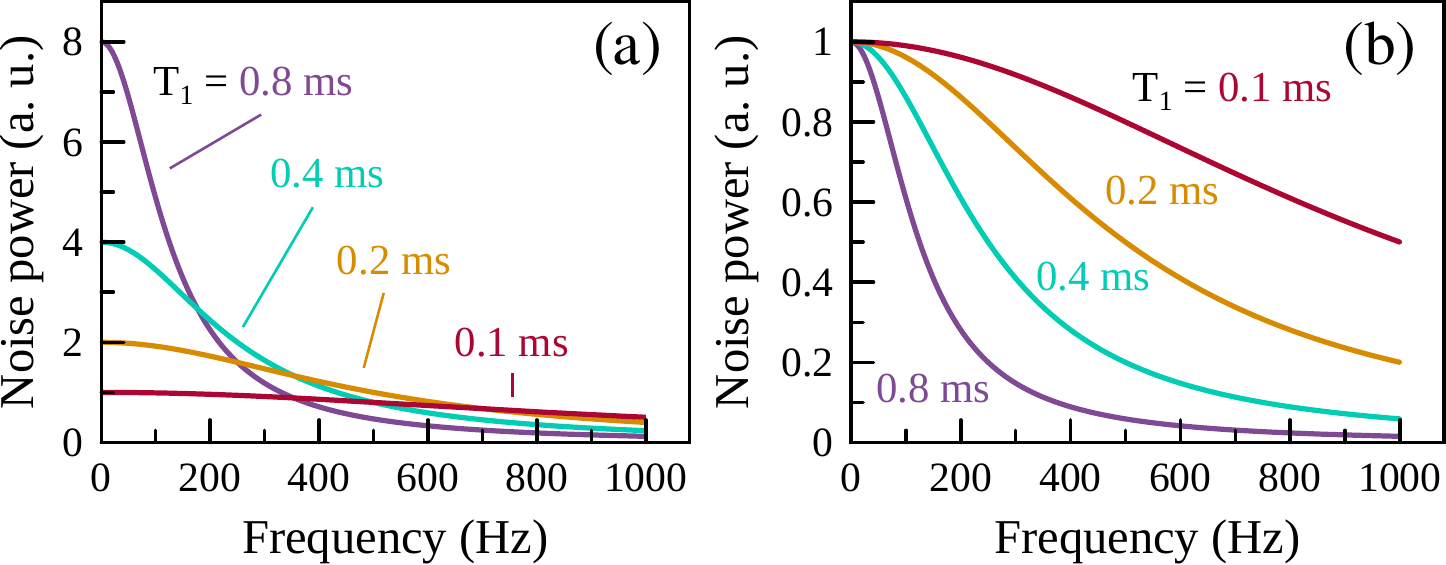}
\caption{Spectra of the longitudinal spontaneous (a) and stimulated (b) spin noise power for different values of the longitudinal relaxation time $T_1$.}
\label{fig1}
\end{figure}

Let us look at spontaneous SN of a system with Hamiltonian~\eqref{1} at $H_x=h=0$. 
In the accepted high-temperature approximation, the mean-square magnetization fluctuation does not depend on the magnetic field or temperature and, in the units of $g^2\beta^2$ equals $1/4$. 
This mean-square fluctuation is equal to the `area' of the noise spectrum~\eqref{5}, which, in this case, has the form 
\begin{equation}
	{\cal N}(\omega)={1\over 4\pi T_1}{1\over 1/T_1^2+\omega^2}, 
	\label{6}
\end{equation}
(here, $\int {\cal N}(\omega)d\omega=1/4$). 
One can see from Eq.~\eqref{6} that when the relaxation time, for some reason, changes, the noise power, at a fixed frequency, may either increase or decrease, passing through a maximum at $T_1=1/\omega$. 
In the spectroscopy of stimulated spin noise, we detect module of the AC susceptibility squared $|\chi (\omega)|^2$, which, in this case, is given by the formula
\begin{equation}
	|\chi(\omega)|^2={\chi_\textit{st}^2\over 1+\omega^2 T_1^2}.
\end{equation}
Herefrom, one can see that, in contrast to the case of spontaneous SN, the stimulated response always decreases with increasing relaxation time $T_1$. 
The difference in the behavior of the noise spectra for these two cases is illustrated by Fig.\ \ref{fig1}. 

The above distinction between dependences of the spontaneous and stimulated SN spectra on the relaxation time $T_1$ indicates a difference between properties of the `hand-made' stochastic perturbation and `natural' perturbation provided by thermal bath of the spin system. 
In the latter case, the intensity of the effective stochastic magnetic field \emph{grows} with increasing relaxation time: it is this growth that provides constancy of the noise power area with its narrowing. 
To maintain the constant area of the SN spectrum (with increasing $T_1$), in the spectroscopy of stimulated noise, one should increase the amplitude of the applied stochastic field. 

Perhaps this curious property of the effective stochastic perturbation in the spectroscopy of spontaneous SN is related to artificial nature of the notion of this kind of perturbation. 
The increase of the spin-system relaxation time indicates a decreasing effect of the thermal bath. 
Under these conditions, the fluctuating motion of the spin system is getting slower, and spectrum of the spontaneous noise narrows, while the magnitude of the fluctuations and the related area of the noise spectrum, in conformity with Eq.~\eqref{6}, remains the same.
  
\section{\label{sec3}Experimental}

\begin{figure}[t]
\includegraphics[width=\columnwidth]{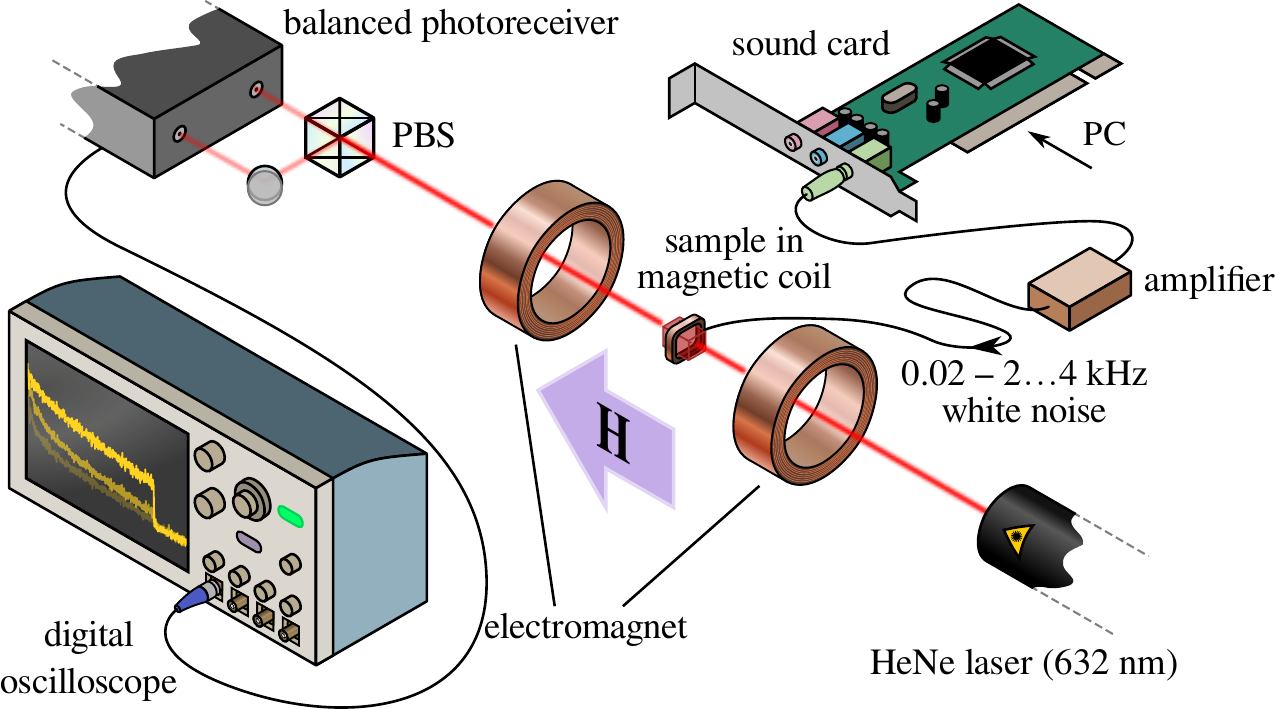}
\caption{Schematic of the experimental setup.}
\label{fig2}
\end{figure}

The measurements were performed at $T \approx 5.5$~K with the sample $\mathrm{BaF}_2:\mathrm{U}^{3+}$ (ground-state configuration 5f$^3$) characterized by strong interconfigurational (5f-6d) absorption and strong magneto-optical activity~\cite{Ant}. 
The concentration of the impurity ions determined from the calibrated absorption spectra presented in Ref.\ \onlinecite{Su} was found to be $\sim 0.08$~mol.\%. 
An additional advantage of this ion was related to zero nuclear-spin of this isotope ($^{238}$U), which made it possible to avoid the strong Van-Vleck contribution usually resulted from hyperfine interaction in the ground electronic state~\cite{Univ}. 
The experimental setup, in many respects, is the same as that for conventional SN measurements (Fig.\ \ref{fig2}). 
As a light source, we used a He-Ne laser with the wavelength ($\lambda = 632.8$~nm) hitting the region of highest Faraday rotation of the crystal \cite{Ant}. 
The sample was placed in a closed-cycle liquid-helium system Montana Cryostation that allowed us to cool the sample down to $5.5$~K in a magnetic field up to $0.1$~T. 
The noise signal, in the spectroscopy of stimulated SN, does not depend on the beam cross section. 
So, the measurements were performed in a collimated light beam. 
An additional fluctuating magnetic field aligned along the DC field was produced by a small coil wound around the sample. 
The frequency spectrum of the fluctuating magnetic field was obtained with the aid of a PC sound card. 

Polarization fluctuations of the light transmitted through the sample were detected in a standard way (using a balanced photoreceiver) and spectrally analyzed with an FFT spectrum analyzer on the basis of a Tektronix digital oscilloscope. 
Accumulation time of a single spectrum usually lied in the range of a few minutes. 

In the simplest measurements, the spectrum of the fluctuating AC field was chosen `white' within the bandwidth of interest for the studied system (usually up to a few kHz). 
At the same time, the proposed experimental approach offered us a highly important additional degree of freedom related to the possibility of arbitrary shaping the spectrum of the applied fluctuating field. 
Specifically, when the frequency resolution of the AC susceptibility spectrum is not needed to be high, the continuous spectrum of the applied fluctuating field can be replaced by a comb-like or some other discrete spectrum (with `white' envelope). 
In this case, one can do with much lower power delivered to the coil (avoiding effects of heating and mechanical vibrations) with no loss of sensitivity of the measurements. 

\begin{figure}[t]
\includegraphics[width=\columnwidth]{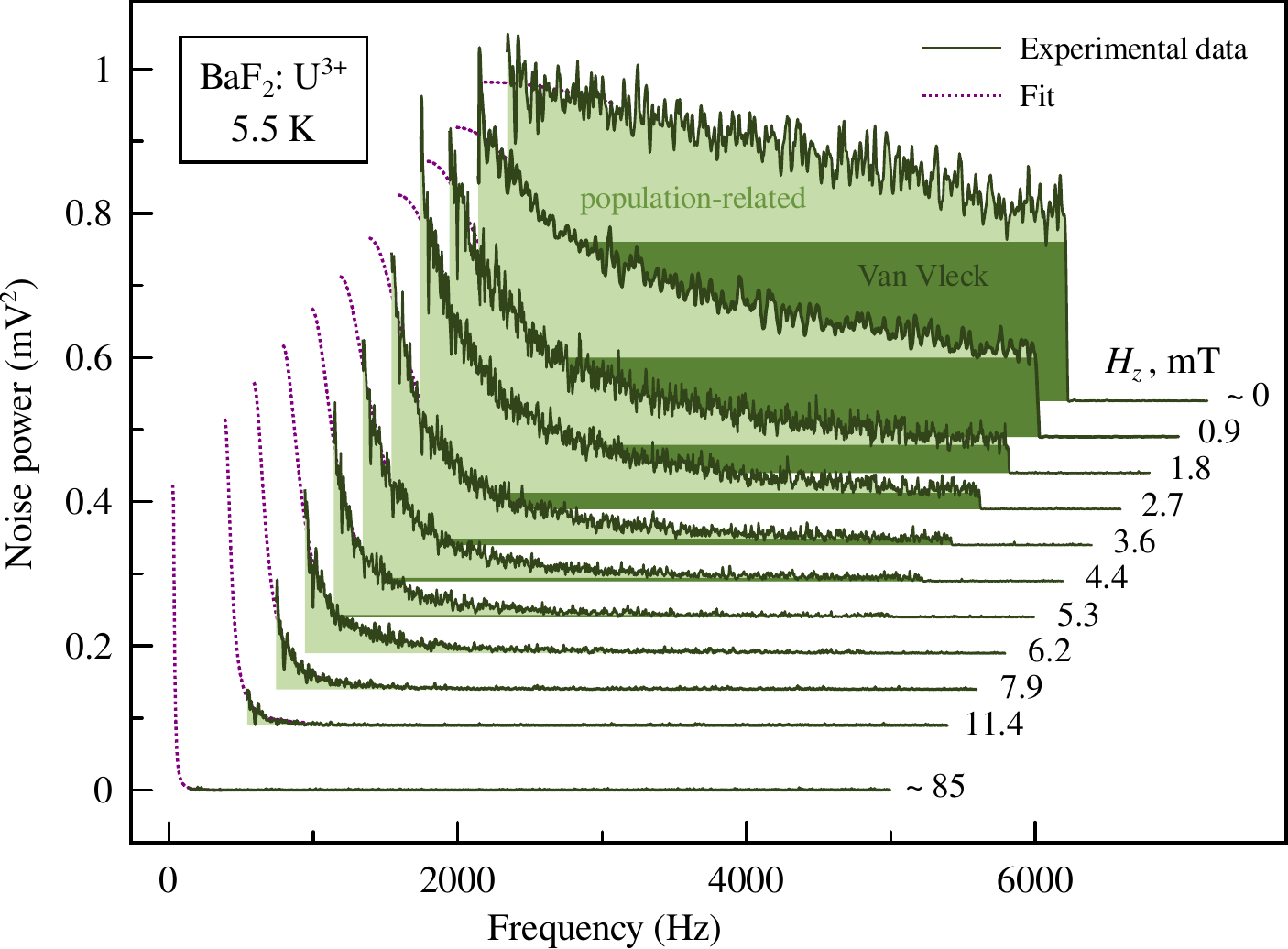}
\caption{Spectra of the stimulated SN power of the $\mathrm{BaF}_2 : \mathrm{U}^{3+}$ crystal at different magnetic fields ($T \approx 5.5$~K). The bright and dark areas correspond, respectively, to the population-related and Van-Vleck contributions.}
\label{fig3}
\end{figure} 
 
\section{\label{sec4}Results of the Measurements}

Feasibility and utility of the proposed method are illustrated by Fig.\ \ref{fig3} where we show experimental spectra of the stimulated SN obtained on the $\mathrm{BaF}_2 : \mathrm{U}^{3+}$ crystal at $T = 5.5$~K for different values of the applied longitudinal magnetic field. 
A spectrum of the fluctuating AC field was flat within the frequency range $0.05$--$4$~kHz and was bounded by this interval. 
As one can see, the SN spectrum, under these conditions, contained, along with the Lorentzian peak characterizing spin relaxation rate of the system, a flat pedestal that decreased with increasing DC field. 
An important feature of the SN signal detected in this way was that it fell down with temperature according to the law $1/T$, in agreement with the Curie law. 
This signified that both components of the SN signal were related to the magnetization of the spin-system, and contribution of the diamagnetic Faraday effect of the crystal host or cryostat windows into the detected noise signal was negligible.

\begin{figure}[t]
\includegraphics[width=0.9\columnwidth]{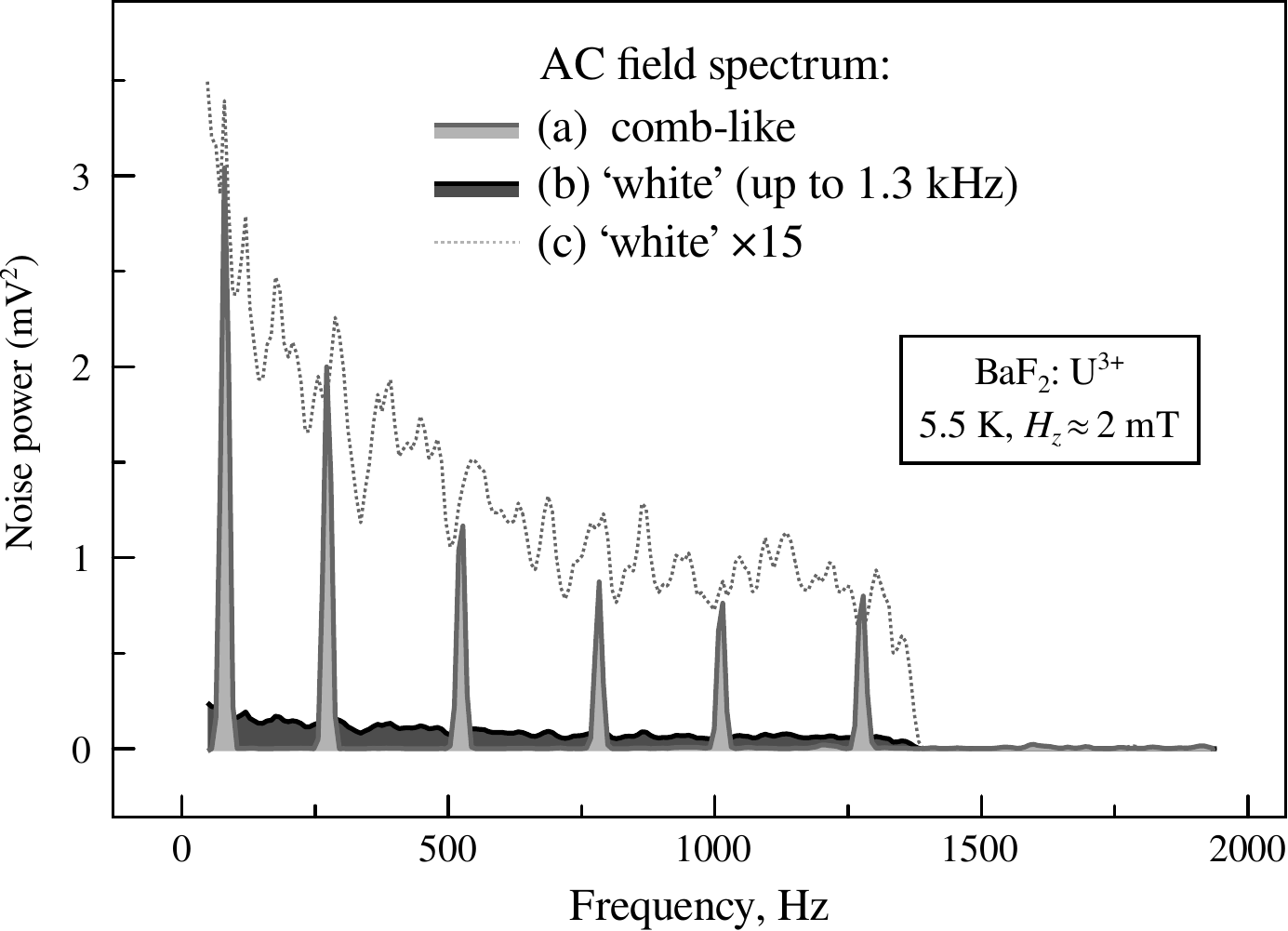}
\caption{Spectra of the stimulated SN power of the $\mathrm{BaF}_2\text{-}\mathrm{U}^{3+}$ for the comb-like (a) and `white' (b) spectrum of the applied fluctuating magnetic field (with equal electric power). For comparison, curve (c) reproduces curve (b) multiplied by a factor of $15$. It is seen that the use of the comb-like spectrum considerably improves the signal-to-noise ratio.}
\label{fig4}
\end{figure}

Figure\ \ref{fig4} demonstrates how the sensitivity of the measurements can be substantially increased using the comb-like spectrum of the applied fluctuating field. 
The figure shows two spectra of stimulated SN for the same power of the AC field delivered to the modulating coil (the same mean-square current), but with its spectrum being either continuous or quasi-discrete. 
One can see that when passing from continuous to quasi-discrete spectrum, the number of meaningful values, within the studied frequency range, substantially decreased, while the signal-to-noise ratio in the remaining points strongly increased. 
 
\section{\label{sec5}Discussion}

The spectrum of the spin noise stimulated by the `white' fluctuating field is expected to provide frequency dependence of the AC susceptibility of the crystal, which, in turn, should include two contributions [see Eq.~\eqref{3}]. 
One of them is controlled by populations of magnetic sublevels and, therefore, responds to the magnetic field variations with a delay governed by the longitudinal relaxation time. 
The other one, mentioned above as adiabatic or of Van-Vleck type, is related to the effects of state mixing and is revealed at small magnetic fields when Zeeman structure of the impurity ion appears to be distorted by a `nondiagonal' perturbation (hyperfine, superhyperfine, and crystal field). 
In Sec.\ \ref{sec2}, this nondiagonal interaction was simulated by the term $\sim H_x$. 
 
\begin{figure}
\includegraphics[width=0.7\columnwidth]{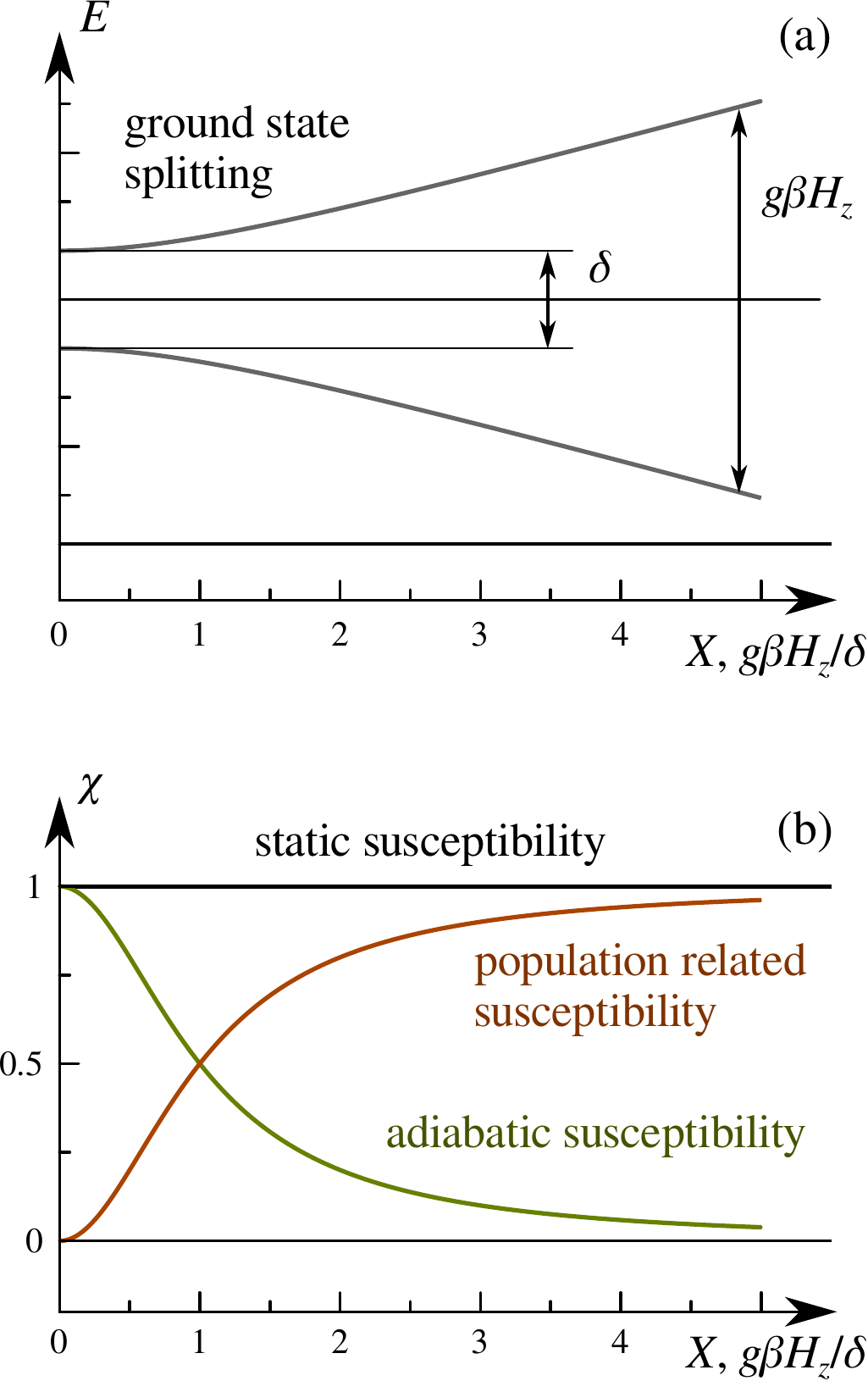}
\caption{ (a) Energy level diagram of the spin system described by Hamiltonian (1) and (b) field dependence of the adiabatic and population-related contributions to the total spin-system susceptibility. }
\label{fig5}
\end{figure}

Schematically, the field dependence of these two contributions at $\omega =0$ can be presented as shown in Fig.\ \ref{fig5}. The spin system ($S_\textit{eff} = 1/2$) perturbed by this nondiagonal interaction acquires a zero-field splitting $\delta=g\beta H_x$. 
For this simplified scheme, the adiabatic contribution decreases with magnetic field $H_z$ as $\chi_\textit{st}H_x^2/[H_x^2+H_z^2]$, while the population-related component increases with the field in such a way that their sum remains constant and corresponds to the static magnetic susceptibility of the system [see Eq.~\eqref{3}]. 
With increasing frequency $\omega$, however, the population-related contribution decreases as $(1 + \omega^2 T_1^2 )^{-1}$, while the adiabatic contribution remains the same.

\begin{figure}
\includegraphics[width=\columnwidth]{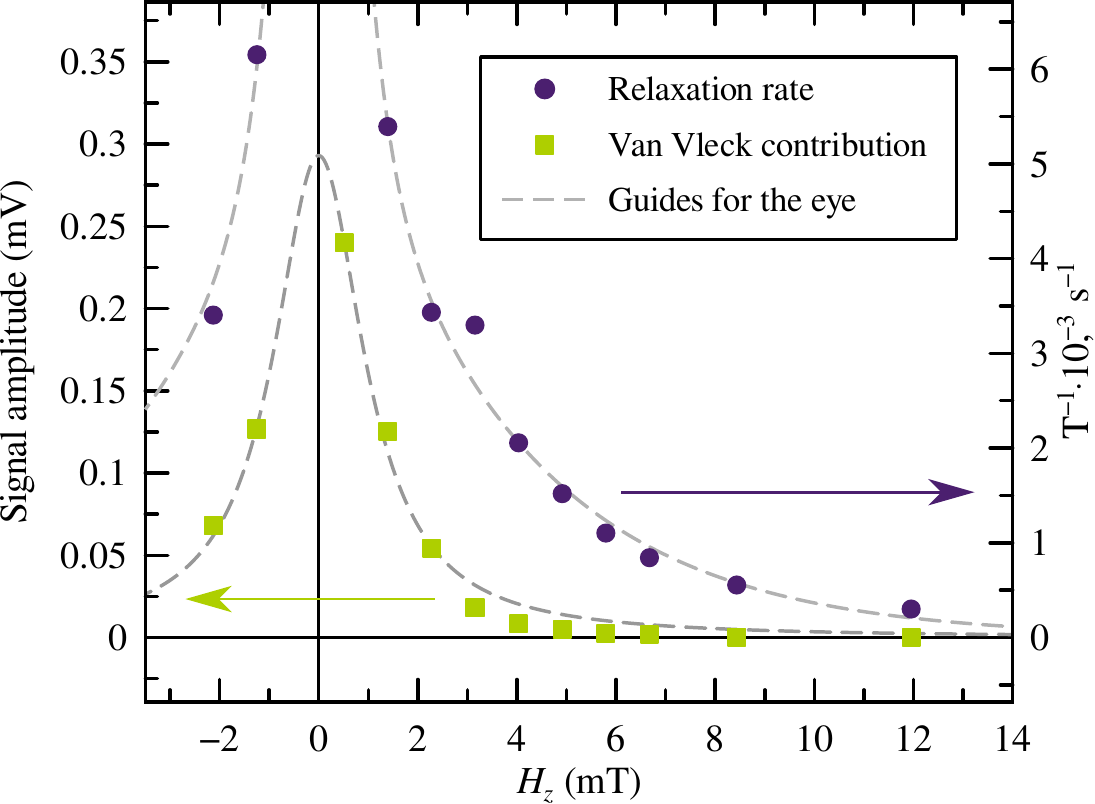}
\caption{Magnetic-field dependence of the longitudinal relaxation rate and Van-Vleck susceptibility of the BaF$_2$:U$^{3+}$ crystal at 6 K. }
\label{fig6}
\end{figure}

Our experimental findings, shown in Fig.\ \ref{fig3}, well correlate with this simplified model. 
Each spectrum of the stimulated SN, obtained in the `single-shot' experiment (Fig.\ \ref{fig3}), allowed us to extract two quantities---longitudinal spin relaxation time $T_1$ and magnitude of the Van-Vleck contribution to the signal at a given magnetic field. 
The former was derived from the width of the population-related Lorentzian, while the latter---from the value of the frequency-independent component of the signal. 
The Van-Vleck contribution naturally decreases with the magnetic field, as the `nondiagonal' perturbation of the spin-system is getting smaller and smaller. 
Figure\ \ref{fig6} shows the magnetic-field dependence of these two quantities.

As seen from this figure, the width of the Van-Vleck peak (vs magnetic field) is rather small ($\sim 2$~mT). 
Similar studies of $\mathrm{CaF}_2\text{-}\mathrm{Tm}^{2+}$ crystals \cite{Zap-mod, Univ} performed using modulation magneto-optical spectroscopy, have revealed two components of the Van-Vleck susceptibility. 
One of them was an order of magnitude broader ($\sim 20$~mT) and was attributed to hyperfine interaction in the ground state of the ion ($S = 1/2$, $I = 1/2$). 
In the case of $\mathrm{U}^{3+}$ ion, with no nuclear spin, this component is evidently absent. 
The width of the second component, which was considered to be a result of superhyperfine interaction, was approximately equal to the one measured here. 
So, we believe that the Van-Vleck peak of AC susceptibility obtained in this work is also caused by perturbation of the $\mathrm{U}^{3+}$ ground state by surrounding nuclear spins of the crystal lattice. 
This type of perturbation makes the energy structure of the impurity system in small magnetic fields much more complicated than that depicted in Fig.\ \ref{fig5}. 
As a result, the spin relaxation rate, under these conditions, decreases with increasing magnetic field rather than increases as it is usual for EPR spectroscopy in higher magnetic fields. 

\section{\label{sec6}Conclusions}

Thus, from the viewpoint of the AC susceptibility, our experimental results look quite consistent, while from the viewpoint of SN spectroscopy, the spectra of spontaneous and stimulated SN, being equally controlled by the AC susceptibility, appear to be essentially different. 
In particular, in the conventional spectra of longitudinal SN, we usually observe the Lorentzian peak at zero frequency with no ``white'' (frequency-independent) background even in small fields. 
The point is that the mean-square fluctuation of the magnetization, in the conventional SN spectroscopy, does not depend on spin relaxation time $T_1$, and, therefore, its peak power at zero frequency increases with $T_1$. 
Correspondingly, with decreasing $T_1$, the noise peak power decreases and for sufficiently fast relaxation becomes negligibly small. 
In the stimulated SN spectroscopy, when the power of the applied fluctuating field is controlled by the experimentalist, and the value of the mean-square fluctuation does not have to be conserved, the AC susceptibility at zero magnetic fields (where $T_1 \rightarrow 0$) does not drop in spite of broad band of response. 

We can conclude that the experimental approach proposed in this paper bridges the gap between the spectroscopy of AC susceptibility and spectroscopy of spin noise. 
This approach inherited from these two spectroscopic methods the idea of linear response and the idea of extracting information from noise. 
As compared with the spectroscopy of spontaneous spin noise, the proposed method does not impose any serious requirements upon the FR cross-section of the spin system, does not imply the use of narrow laser beams, and, in principle, does not care much about polarimetric sensitivity of the setup, because magnitude of the signal is controlled by the external fluctuating magnetic field. 
In our opinion, this method may find practical application in magnetic spectroscopy of crystals and glasses with paramagnetic impurities. 

\section*{Acknowledgements}
The authors collaborative team highly appreciate the support from RFBR-DFG grant No.\ 19-52-12054.
We acknowledge the financial support by the Deutsche Forschungsgemeinschaft in the frame of the International Collaborative Research Center TRR 160 (Project A5).
The authors from Russian side acknowledge Saint-Petersburg\ State\ University for a research grant ID\ 40847559.
The work was fulfilled using the equipment of the SPbSU Resource Center ``Nanophotonics''.

\end{document}